\documentclass[11pt]{article}
\usepackage{amsfonts}
\usepackage{amssymb}
\usepackage{amstext}
\usepackage{amsmath}
\usepackage{xspace}
\usepackage{theorem}
\usepackage{graphicx}
\usepackage{url}
\usepackage{graphics}
\usepackage{colordvi}
\usepackage{colordvi}
\usepackage{subfigure}
\usepackage{lineno,hyperref}
\modulolinenumbers[5]
\usepackage{verbatim}
\usepackage{tabularx}
\usepackage{float}
\usepackage{placeins}
\usepackage[authoryear]{natbib}
\usepackage{amssymb}

\title{Using text mining and machine learning for detection of child abuse}

\author{Chintan Amrit\thanks{IEBIS Department, University of Twente, 7500 AE, Enschede Netherlands. Email: {\tt c.amrit@utwente.nl}}\and Tim Paauw\thanks{Ynformed, The Netherlands. Email: {\tt timpaauw@gmail.com}} \and Robin Aly\thanks{Database group, University of Twente, 7500 AE, Enschede Netherlands.  Email: {\tt r.aly@utwente.nl}}. \and Miha Lavric \thanks{IEBIS Department, University of Twente, 7500 AE, Enschede Netherlands. Email: {\tt m.lavric@utwente.nl}}}

\begin{document}
\begin{titlepage}
\maketitle

\thispagestyle{empty}

\begin{abstract}
Abuse in any form is a grave threat to a child's health. Public health institutions in the Netherlands try to identify and prevent different kinds of abuse, and building a decision support system can help such institutions achieve this goal. Such decision support relies on the analysis of relevant child health data. A significant part of the medical data that the institutions have on children is unstructured, and in the form of free text notes. In this research, we employ machine learning and text mining techniques to detect patterns of possible child abuse in the data. The resulting model achieves a high score in classifying cases of possible abuse. We then describe our implementation of the decision support API at a municipality in the Netherlands. 

\end{abstract}

\end{titlepage}

\section{Introduction}
\label{sec:introduction}

Child abuse is serious problem, with an estimated 40 million children being subject to abuse or neglect each year ~\citep{world2001prevention}. 
For 2014 alone, there have been 3.62 million referrals of alleged child maltreatment in the USA,  
resulting in the identification of 702000 victims (including 1580 fatalities) of child abuse and neglect. Despite these high numbers of registrations and identified victims, cases of child abuse  still remain unregistered and/or unidentified, due to missing and incomplete information, preventing adequate procedures ~\citep{Gov2014}. In the same year, an UK inquiry into child sexual abuse in the family environment by the Children's Commissioner showed that as little as only 1 in 8 victims of sexual abuse may come to the attention of the police and children's
services, with up to two thirds of all sexual abuse happening in and around the family~\citep{UK2015}. 
One way to improve the registration of child abuse is by providing training to stakeholders. Indeed, a German study concluded that everyone working in the area of child protection should receive additional interdisciplinary training ~\citep{bressem2016german}. However, such training might prove to be costly and time consuming.

An alternative approach is to provide child health care professionals with a decision support system, assisting them to identify cases of possible child abuse with a higher precision and accuracy. Recent research has tried to predict cases of child abuse using structured data \citep{gillingham2015predictive, horikawa2016development}. While these methods achieve a reasonable performance, they do not take the knowledge of the pediatrician into account \citep{goldman1990tacit}. One of the sources of evidence that health care professionals often create as part of their daily procedures, is free-text. As such texts are less constraining than structured data, they possibly incorporate elements of doctors' tacit knowledge of the phenomena that are not included in structured data \citep{malterud2001art, henry2006recognizing}. 
In this article we propose a decision support based approach to increase the number of correctly identified child abuse cases and improve their registration. We aim at providing health care professionals timely and appropriate decision support about possible child abuse based on patterns in health data that the health care professionals create as part of their daily procedures. 

The findings of this article are validated based on data from the Netherlands, where each child visits the public health organization (GGD\footnote{GGD from its Dutch spelling: Geneeskundige en Gezondheidsdienst.}) roughly 15 times between the ages zero and four. The pediatricians and nurses performing these consultations use information systems to keep track of each child's development. Depending on the type of consult, both structured and unstructured data are recorded, where structured data can be the child's height or weight and unstructured data consists of free-text containing the pediatrician's remarks during the consultation. We first explore whether these consultation data contain meaningful patterns concerning child abuse. We then investigate whether machine learning from this data can help in identifying cases of child abuse. We therefore train our machine learning classifiers on cases of abuse as determined by over 500 child specialists from a municipality of The Netherlands. We perform a methodological evaluation of a wide range of methods to identify their strengths and weaknesses. We then evaluate the automatic classifications with judgments of pediatricians, and thereby address our intention to provide decision support (in identifying child abuse) for pediatricians. We finally describe our implementation of the decision support API at the municipality. 

In summary, our contributions are the following:
\begin{enumerate}
\item A study to determine, if pediatrician data contains patterns indicating child abuse,
\item A comparison of machine learning methods for this task, 
\item The evaluation of some of these methods through health care professionals, and
\item An implementation of the prediction model in the municipality through an API.
\end{enumerate}

The remainder of this paper is structured as follows: Section 2 describes related work, Section 3 presents the models and methods we utilized in our research. Section 4 describes how we pre-processed the data, Section 5 describes the results, section 6 discusses the results and the API implementation and finally section 7 describes the conclusions and future work.




\section{Related work}
\label{sec:relatedwork}

This paper's contribution is related to work in the area of data exploration and supervised classification based on machine learning in the medical texts. 

Previous work has been done in the overlapping fields of medical data mining, medical NLP or BioNLP and medical text mining \citep{chapman_current_2009, vanderSpoel2012}. Relevant to this research are studies that focus on data mining or text mining in the (semi-) medical context. Closest to this research are applications of predictive analytics using unstructured (semi-) medical text.

\subsection{Data mining in the medical context}
\label{sec:rw_dm}
 \cite{bellazzi_predictive_2008} provide an overview of data mining in clinical medicine and propose a framework for coping with any problems of constructing, assessing and exploiting models. The emphasis is mainly on data mining in general, but useful guidelines are provided. \cite{yoo_data_2011} provided a similar literature review more recently.

Apart from research by \cite{rao_predicting_2012} towards classification of breast- or bottle-feeding from unstructured data, no previous research uses unstructured semi-medical data for predictive analysis. There is no precedent of identifying child abuse from semi-medical texts.

\subsection{Text mining in the medical context}
\label{sec:rw_tm}
Many text mining studies in the medical context are focused on extracting structured knowledge from medical text or notes. Efforts towards creating a pipeline for analysing medical notes include the work by \cite{goryachev_suite_2006}, who define three typical uses of such pipeline: to match concepts, to construct a classification model, or to automatically encode the documents using for example Unified Medical Language System (UMLS) or Medical Language Extraction and Encoding System (MedLEE). \cite{zeng_extracting_2006} then use the pipeline to extract data related to asthma from free text medical notes. Automatic encoding makes up a significant part of the available literature, including work by  \cite{friedman_automated_2004}, \cite{zhou_approaches_2006} and  \cite{hyun_exploring_2009}. Other efforts include extraction of  disease status from clinical discharge texts by \cite{yang_text_2009}. An overview of such research with the aim of supporting clinical decisions is given by \cite{demner-fushman_what_2009}.

Most of the developed tools and methods involve the English language, but there are a few occasions in which Dutch medical language was studied. \cite{spyns_dutch_1996} pioneered with a Dutch Medical Language Processor (DMLP) focusing on the language-specific parts of the language processing chain or pipeline. They later evaluated their work with four applications, concluding that although work still had to be done, the results were very promising \citep{spyns_dutch_1998}. Up to that point, an overview by \cite{spyns_natural_1996} shows that only one attempt at processing Dutch clinical language had been done with the M{\'e}n{\'e}las project \citep{zweigenbaum_coding_1995}. This project was mainly a free-text encoding effort.

More recently, \cite{cornet_inventory_2012} provide an overview of the three tools available for Dutch clinical language processing with the goal of outputting Systematised Nomenclature of Medicine Clinical Terms (SNOMED CT) data. SNOMED CT is the most comprehensive collection of systematically organized medical terms in the world. It is multilingual and is used mainly to effectively record and encode clinical data. \cite{cornet_inventory_2012} also emphasize that a lot of research is needed towards a spelling checker, a negation detector and importantly a domain-specific acronym/abbreviation list as well as a concept mapper for the Dutch medical language. In summary, after the research by \cite{spyns_natural_1996} and \cite{spyns_dutch_1996,spyns_dutch_1998}, not much progress has been made over the last 20 years for Dutch medical language processing.

While this research is focused on machine learning thus automatic pattern extraction, it is interesting to review literature for indicators abuse, which in contrast to obesity, is hard to capture in structured data. These insight can later on be used for selection of structured data. Structured data can be used to improve the model, and to test whether analyzing the unstructured data is even needed next to the structured data.

\subsection{Child abuse}
\label{subsec:litabuse}
A comprehensive Delphi study\footnote{A Delphi study is a research method that relies on a panel of experts. In several rounds, experts are asked for their opinion on a subject of disagreement. After each round, an anonymous summary of all experts judgments is provided and experts are encouraged to adjust their opinion in the light of this summary. Iteratively, consensus is to be achieved.} by \cite{powell_early_2003} on early indicators of child abuse and neglect describes 5 physical (e.g. patterns of injuries), 13 behavioral/developmental indicators (e.g. self-harm, undue fear of adults) and 16 parenting (e.g. inability to meet basic needs of child, use of excessive punishment) indicators, that may occur separately or cluster together. 
Although the indicators of child abuse are an ambiguous topic, this Delphi study among 170 experts from different backgrounds does provide an interesting overview. The author herself points at a  a potential flaw of the study, as most of the experts on the panel held senior positions in their organizations and did not reflect the first-line child protection. Moreover, in a generally favorable commentary of the review \citep{sidebotham2003red}, it has been stated that whichever (clusters of) indicators are used to alert people to possible maltreatment, they are not diagnostic and definitive proof of maltreatment, and that taking the step from  a possible indication to a diagnosis of maltreatment requires clinical acumen and a holistic approach.
A study towards the demographics of abuse was conducted by \cite{jones_links_1992} identifying links with types of maltreatment. Unfortunately, the test group in their study consists solely of abused children and thus no relevant indicators of abuse versus no abuse were identified.

\subsection{Predicting child abuse}
\label{subsec:litpredabuse}
More than 30 years ago the development of expert systems has been proposed for providing decision support to professionals in child protection agencies \citep{schoech1985expert}. This is becoming a reality in today's environment, where a substantial amount of data from multiple sources is available on children and their families \citep{gillingham2015predictive}. Building upon the view of \cite{schoech2010interoperability}, in an ideal scenario the gathered information should not be modified by the use of an expert system, while at the same time allowing potential modification of the existing patterns of information flow to be more efficient. Moreover, for child welfare workers this would mean good interoperability in a top-down model, where the expert system could assist with all manner of tasks, ranging from routine/inconsequential (e.g. recording information) up to those of critical importance, providing support with assessment of risks pointing towards child maltreatment.

Predictive risk modeling (PRM) tools coupled to data mining with machine-learning algorithms should be capable to direct early interventions to prevent child maltreatment from occurring \citep{gillingham2015predictive}. Successful early intervention programs already exist e.g. the Early Start program in New Zealand \citep{fergusson2006randomized}, however there is a range of challenges that need to be addressed before coupling these programs with PRM. In addition to selecting reliable and valid outcome variables, while ensuring the consistency of their registration \citep{gillingham2015predictive}, there are moral and ethical challenges that need to be taken into consideration \citep{keddell2014ethics}.

\cite{vaithianathan2013children} explored the potential use of administrative data for targeting prevention and early intervention services to children and families. Their data set was derived from public benefit and child protection records from the 57,986 children born in New Zealand between January 2003 and June 2006 and recorded until 2012. The final predictive risk model, with an area under ROC curve of 76\%, included 132 variables. From the top 10\% children at risk, 47.8\% had been substantiated for maltreatment by age 5 years. Of all children substantiated for maltreatment by age 5 years, 83\% had been enrolled in the public benefit system before the age of 2.

\cite{horikawa2016development} developed a linear prediction model (45,2\% sensitivity, 82,4\% specificity) using administrative data from 716 child maltreatment incident cases (stringently selected from 4201 cases reported to Shiga Central Child Guidance System, Japan) to identify the first recurrence of child abuse within the first year of the initial report. They identified and used 6 factors in their multivariate logistic regression model,  namely the age of child, the age of the offender, the history of abuse of the offender, household financial instability/poverty, absence of guardianship and referral source.
\section{Models and methodology}
\label{sec:models}

\label{subsec:datacollection}
For this research, data was provided from the child health department (JGZ) of the largest public health organization in the Netherlands, the GGD Amsterdam. In addition, JGZ also provided knowledge and expertise in the form of pediatricians in a scrum group. 
The data consisted of (partly medical) files on 13.170 children born in 2010 in the Amsterdam region, all reaching the age of four in 2015, at time of this research. With on average 14,8 contacts with the JGZ per child, these visits resulted in 195.188 individual data entries.  Of the 13.170 children, 657 children had been labeled presumably abused by the JGZ over the course of four years. It is important to note that the JGZ estimated that these 657 children account for 25\% to 30\% of children that should have been labeled. An overview of the data's layout is given in ~\ref{app:initiallayout}.
%
%
\subsection{Data Exploration}
Quantitative characteristics of the data set are summarised in \autoref{tab:datachar}. Taking all children born in one year ensures relative randomness of the sample. The year 2010 was chosen, as the current information system Kidos and JGZ way of working were already in place and established, providing a stable environment for data retrieval, not needing additional data transforming steps between systems.  
With regard to privacy, any structured information that could be used to identify the child was removed, e.g. a unique identifier per child added by the JGZ to enable tying pieces of data together for one child. References to staff were handled in a similar manner. As described by \cite{cios_uniqueness_2002}, this process of \textit{de-identification} ensures anonymity but allows for the JGZ to trace back specific results to specific children.

%
%
\begin{table}[!htbp]
\caption{Characteristics of the GGD data set}
\label{tab:datachar}       
\def\arraystretch{1}
\begin{tabularx}{\textwidth}{Xl}
\hline\noalign{\smallskip}
\textbf{Characteristic} & \textbf{Value}  \\
\noalign{\smallskip}\hline\noalign{\smallskip}
Number of children & 13.170 \\
Consults & 195.188 \\
Average number of consults per child & 14,82 \\
Average number of words per consult & 41,58 \\
Lexical diversity (nr of unique words vs. total nr of words) 1k random consults & 0,16 \\
Lexical diversity 1k random consults excluding stopwords & 0,23 \\
\noalign{\smallskip}\hline
\end{tabularx}
\end{table}

\subsection{Unstructured data}

The data used in this research are the (semi-)medical notes written down by pediatricians or nurses into four, subject specific fields for note-taking per consult, the most voluminous being the \textit{conclusion} field. This field contains a summary of the child and is hereafter referred to as SOC: summary of child. 
Some of the text is about the social dynamic of the family,  describing the current situation, wishes of the parents and a number of medical diagnostics. The text contains numerous acronyms depending on the author and the team the author is part of. The average amount of words per consult is 41.58.

An example of a short note taken 4 months after birth of another child is:
\begin{quote}
\textit{prima kind, m chron bronchitis advies begin fruit pas met 5 mnd}
\end{quote}
Translating to English as follows:
\begin{quote}
\textit{Nice child, mother has chronic bronchitis, advised to not start with fruit until 5 months of age.}
\end{quote}

\subsection{Structured data and labels}
\label{sec:struc_data}
In addition to the unstructured notes, we added specific structured data and labels 
to our data set 
that we used as features, and as dependent variables in our predictive model. These data/labels are i) "Findings ZSL", ii) "Action ZSL", iii) "Attention", iv) "Family relations" and v) BMI.    

"Findings ZSL" represent worries in the social environment (Dutch - "Zorgen Sociaal Leefmilieu"), set to 1 if there has been a presumption of child abuse and 0 otherwise. In the analyzed data set, 628 out of 13,170 distinct children had this label set to 1. When we found that 'Findings ZSL' was wrongly set to 0, although nature of abuse was known, a corrective action was taken to include those children as well, amounting to 657 children additionally labeled as presumably abused. 
JGZ indicates that professionals should always set this to 1 for children that they presume to be abused, making "Findings ZSL" in principle useful  as a dependent variable for a predictive model for child abuse. In reality, this happens only in about 25\%-30\% of the cases, either to prevent the risk of drawing a wrong conclusion and hurting the bond of trust with the parents, or because the health professional takes action without registering it. This leads to noise in the data, to incorrect management information and more concretely to missing abused children in the data.  

The "Attention" label is set to 1, if, due to any reason, (extra) attention needs to be paid to this child, with no clear directive given. 2,459 out of 13,170 children, approximately one in five, have this label set to 1.

"Family relations" are summarized in a table, containing relation types (e.g. brother, mother, adoptive father etc.) and ID (birth date only) of the relative.

Also, when a child visits for a consultation, he or she is measured and weighted, resulting in tables of lengths, weights and Body Mass Indexes (BMI). 

\section{Pre-processing}
%
%

\subsection{Storage}
Before the data can be processed, we inserted the data into a MySQL\footnote{\url{http://www.mysql.org}} database, allowing for easy filtering and drilling down on dimensions. With the prospect of engineering features per child, there are multiple fact tables containing for example the \textit{summaries of child} (SOC). The data for distinct children are stored in a dimension table. This way, features like the number of consults or whether or not a child is obese can be extracted to one flat feature table. Other fact tables include the Body Mass Indices (BMI), ZSL and Attentions linked to the children. Other dimension tables include the action types, the locations of the consults and the practicing pediatricians. 

\subsection{Terminology normalization}
\label{sec:termnorm}
There are many abbreviations and acronyms used in the texts that enabled quick input of data, most of these were imposed standards among the JGZ personnel.  Acronyms such as \textit{P} for \textit{papa} translating to 'father' in English and \textit{ZH} for \textit{ziekenhuis} translating to 'hospital' in English. We used  regular expressions (RE) to extract all short abbreviation-like words can be extracted from the data, e.g. words consisting of less that 4 characters that are all consonants and possibly contain dots. This is in agreement with the  method utilized by \cite{xu_study_2007}. The extracted acronyms and abbreviations are then ordered on frequency of appearance and enriched with a sentence in which the acronym appears using the NLTK \textit{concordance} function.

Next, we asked the subject experts, in this case the medical staff of the JGZ, to explain the list of acronyms. We then converted the terms to regular expressions and formed tuples with their respective replacement. We used a Python script  to loop through the data for RE-based string replacements.

\subsection{Trivial word removal}
We removed all indications of time for two reasons; they did not contribute to the identification of abuse and they varied constantly introducing noise in the data. Similarly dates and times were both removed using RE. Finally all left-over numbers or words containing numbers were removed.

Another common preprocessing step is the removal of stop words; words that appear very frequently but do not attribute to the meaning of the text. For this, we used a standard Dutch language stop word list, which is included in the Natural Language Toolkit (NLTK) package. Amongst the stopwords were a number of negative words like "niet" and "geen", or not and no/none in English, that were not removed from the text. This was done to make sure that when n-grams with $ n > 1 $ are used, the meaning of the text \textit{"not good"} is captured.  

\subsection{Stemming}
Words in the SOCs appear in many forms and tenses, whilst pointing to the same concepts. These various forms lead to a more flat distribution of word quantity: more unique words and less volume for the top terms. This is not beneficial for classification algorithms that need to identify common terms and themes. In order to group various forms of the same words together, all words were reduced to their stemmed form using the Dutch Snowball stemmer\footnote{\url{http://snowball.tartarus.org}}. This stemming framework proposed by \cite{porter_algorithm_1980}  is included in the Python NLTK package.


\subsection{Tokenization}
Depending on the method of classification, the texts needed to be split up into sentences or just sequences of words. Although splitting sentences seems like an easy task, it is very hard to perform algorithmically. A sentence might end with any of \textit{'.!?'} followed by whitespace and a (capital) character, but not only do quick notes often not comply with this rule, also this combination of characters is frequently seen in abbreviations or medical terminology. We therefore used \textit{tokenization} to split the text to words. This also allowed us to use n-grams, combinations of \textit{n} sequential words, in the analysis later on. Because all noise regarding line-breaks, special characters and white-space had been removed with RE, tokenization was easily done by splitting the text using the single white-space character. 

\subsection{Clustering}
To explore the data, we clustered the children's data according to the proximity of their feature vectors. Therefore, the SOCs per child were concatenated and represented in a feature vector using a tf-idf weighting scheme. Next, we used K-means clustering, which is one of the most effective although simple clustering algorithms. 
By varying the number of clusters (and by using the elbow method) we tried to determine what were the important topics covered by the text. We then used these clusters for creating the predictive models to assign new instances to. As such, one could imagine picking any of the cluster's themes for predictive analysis. As the clusters become smaller though, it will most likely become harder to get a set of training data from the cluster large enough for good classifier performance.

\subsection{Extraction of possible textual features}
In order to explore possible features in the SOCs for predictive analysis, we used \textit{force} specific clusters and found the most common words or combinations of words. This was done for the whole body of texts and for specific groups of children. In this way, we could uncover words that were features for identifying groups of children, and could this could help us identify the groups that suffer from abuse. We visualised the words and their relative frequency using word clouds and discussed the results with the JGZ, to demonstrate the distinctiveness of groups within the population of children. These resulting word clouds showed some obvious and some interesting terms as being distinctive for a group. This indicated that the topics that should be predicted were described in the SOCs, and thus the SOCs could be usable for further text mining.

\subsection{Extraction of possible summarising features}
For free text data, features can be extracted from the content of the data as well as from the form of the data. The latter are called summarising features and can be very relevant for the performance of the classification model. To illustrate this, we explored two summarising features for child abuse.

\begin{figure}[h]
  \centering
    \includegraphics[width=1\textwidth]{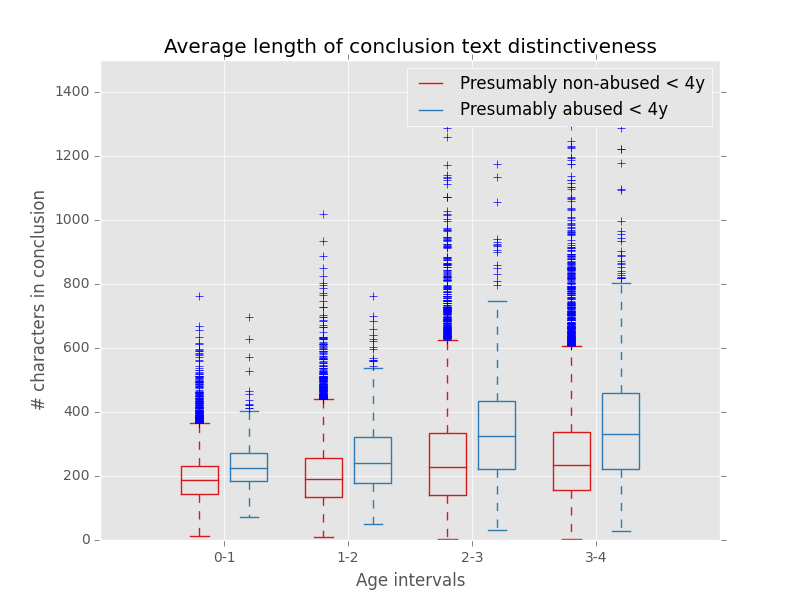}
  \caption{Box plot of the average SOC lengths per age interval for presumed abuse}
  \label{fig:boxplotzsl}
\end{figure}

\paragraph{SOC length as predictor for presumed abuse}
A typical feature that summarises free text is the length of the text. The reasoning that leads to this feature is that more extensive documentation could be made for children that have some health issues (like abuse). We tested if the length of the text (SOC length) could be used as a predictor for presumed abuse. The groups of children were split using the enhanced ZSL finding column calculated per age interval of 0-1, 1-2, 2-3, 3-4. These age intervals were used because we wanted to test whether the SOC length differs significantly between the groups, and from what age could the difference be significant.  \autoref{fig:boxplotzsl} indicates how distinctive the length of the consult really is.

\begin{table}[h]
\caption{Mann-Whitney U-test p-values for average SOC length per age interval for presumed abuse}
\label{tab:pvalueszsl}       
  \def\arraystretch{1}
\begin{tabularx}{\textwidth}{XXX}
\hline\noalign{\smallskip}
\textbf{Interval} & \textbf{Test statistic} & \textbf{p-value}  \\
\noalign{\smallskip}\hline\noalign{\smallskip}
0-1 & 2262729.0 & 1.50948e-45 \\
1-2	& 2253800.0 & 1.26391e-34 \\
2-3	& 1843716.0	& 4.18773e-39 \\
3-4	& 1888747.5	& 3.78134e-39 \\
\noalign{\smallskip}\hline
\end{tabularx}
\end{table}

The p-values in \autoref{tab:pvalueszsl} show that the difference in SOC length between the  presumably abused children and the other children, is significant. Therefore, the average length of the SOC can be used as feature in a predictive model for presumed abuse. It makes sense that the difference between mean values rises over the years, for at the end of year 4, all children that are presumably been abused between 0 and 4 have the label \textit{ZSL finding}. In contrast, between the ages 0 and 1 we would expect about 25\% of children that end up with a \textit{ZSL finding}, already have such a label.

\begin{figure}[h]
  \centering
    \includegraphics[width=1\textwidth]{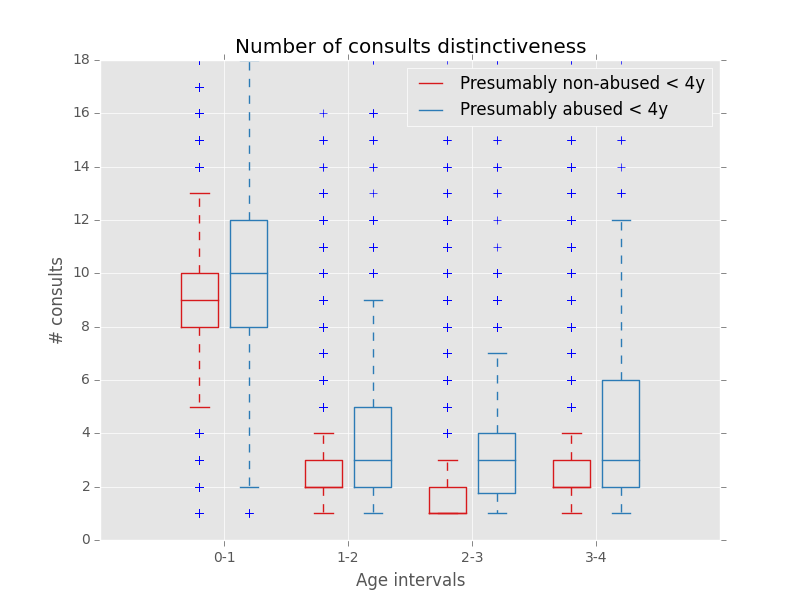}
  \caption{Box plot of the consult quantity per age interval for presumed abuse}
  \label{fig:boxplotzslquantity}
\end{figure}

\paragraph{Consult quantity as predictor for presumed abuse}
The number of consults can also possibly be used as feature in a predictive model for presumed abuse. We used the same groups as previously, using the \textit{ZSL finding} variable. The resulting box plot indicating a difference in average number of consults can be seen in \autoref{fig:boxplotzslquantity}. Again the p-values in \autoref{tab:pvaluescqzsl} indicate that the differences are significant and that this summarising feature can be used as a predictor for presumed abuse.

\begin{table}[h]
\caption{Mann-Whitney U-test p-values for consult quantity per age interval for presumed abuse}
\label{tab:pvaluescqzsl}       
  \def\arraystretch{1}
\begin{tabularx}{\textwidth}{XXX}
\hline\noalign{\smallskip}
\textbf{Interval} & \textbf{Test statistic} & \textbf{p-value}  \\
\noalign{\smallskip}\hline\noalign{\smallskip}
0-1 & 2575126.5 & 2.67548e-25 \\
1-2	& 2110410.5 & 2.76681e-45 \\
2-3	& 1517366.0	& 2.48735e-71 \\
3-4	& 1871244.5	& 1.22809e-40 \\
\noalign{\smallskip}\hline
\end{tabularx}
\end{table}

Other summarising features include the lexical diversity of the SOCs, the average time span between consults or the number of distinct medical professionals a child has come in contact with.

From the previous sections it is evident that the data contains textual and summarising features that could be relevant in the creation of a prediction model for presumed child abuse. One of the contributions of this research is the use of unstructured JGZ data in predicting presumed child abuse, but the main goal of this research was to have a usable prediction model. The JGZ had little structured data available that they had linked to suspicion of abuse. In the next section, we describe our model using structured data and compare it with the one build using unstructured data.

\subsection{Data sampling}
\label{subsec:datasampling}

With just 5\% of the data belonging to the positive group, the data was relatively unbalanced. To be able to use the data for classification modeling, we used the  \textit{random under-sampling} \citep{he_learning_2009} approach. To ensure a large enough training set to cover the various types of abuse, our training data consisted of half of the positive group and an equal number of files from the negative group. We sampled both at random from the positive and negative groups respectively. k-fold cross validation is usually limited to two folds when taking half of the positive group for training. By repetitively sampling the data randomly, cross validation was possible with more than 2-folds.

\subsubsection{Term weighting}
\label{subsubsec:termreduction}
We tested several weighting schemes for possible improvement of the classifier performance: Boolean occurrence, count, tf-idf augmented for varying text lengths \citep{manning_introduction_2008}, DeltaTF-IDF \citep{martineau2009delta} and BM25 \citep{robertson_okapi_1999}.

\subsection{Classification models}
\label{subsec:classmodels}
We used the three most popular algorithms \citep{aggarwal_survey_2012} for this classification task: Naive Bayes \citep{kononenko_inductive_1993}, Random Forest \citep{ho_random_1995} and Support Vector Machine \citep{drucker_support_1999}. Although non-linear algorithms like Neural Networks have received much attention lately, we did not consider them, due to their computational heaviness and history of not consistently outperforming less sophisticated methods in text mining \citep{aggarwal_survey_2012}. We employed the Python implementations of the algorithms, as provided in the widely used Scikit-learn package \citep{pedregosa_scikit-learn:_2011}. We used the same algorithms to classify the structured and unstructured data because of their flexibility and ability to cope with a sparse, high dimensional feature space \citep{aggarwal_survey_2012}, that is typical for text mining.

\section{Evaluation}
\label{sec:evaluation}
After the exploratory analysis, we endeavor to predict whether a child suffers from abuse using classification models. 

\subsection{Performance metrics}
It is important to consider the performance metrics before modelling, for these metrics dictate when a model is performing well. Typically there are many trade-offs between these metrics when optimising a model: improving the model on one metric will decrease the score on another. 

In our model, True Positive (TP) implies a correct classification of an abuse presumption. False Positive (FP) are children that are classified as presumably abused but are not labelled as such by the JGZ. False Negative (FN) are children that have been labelled by the JGZ but are not classified as such by the algorithm. Lastly, True Negative (TN) are correctly classified children that have not been labelled presumably abused. 

The model will be used to find children with a condition, so focusing on recall is important, for as few FN as possible should be predicted. The balance between these two goals is captured in the Receiver Operating Characteristic (ROC) curve and its summarising metric Area-Under-Curve (AUC). Optimising the ROC curve and AUC will be the directive for optimisation of the model. To prevent the classification model from assigning every child a presumably mistreated model, accuracy and balanced accuracy should also be taken into account, even more so since the classes are very unbalanced ($p' = 657$ versus $n' = 13,137$). 

As we described in \autoref{sec:struc_data}, there was a lot of noise in the dependent variable \textit{ZSL finding}, that is our dependent variable of our model. The JGZ indicates that for every child who has a correct ZSL label, around 3-4 children would have an incorrect label. It is therefore unclear if patterns indicating abuse will be found not only in the group with the label ZSL, but even more so in the in the non-ZSL group. As a result, the amount of FP will always be high and the precision low. Indeed, if the precision would be optimised too much by lowering the amount of FP, the goal of spotting children that might be abused is not achieved. We can therefore expect the precision to be low due to the noise in the data, which is precisely the problem that we aim to solve by deploying this model for decision support. Thus, though we mention precision scores in the optimisation tests they are not so critical in determining the quality of our model.


\subsection{Data selection}
We assigned the value 1 for presumably mistreated (ZSL) if at any point the particular child had been labelled so by JGZ, and 0 if not.

\begin{figure}[h]
    \includegraphics[width=1\textwidth]{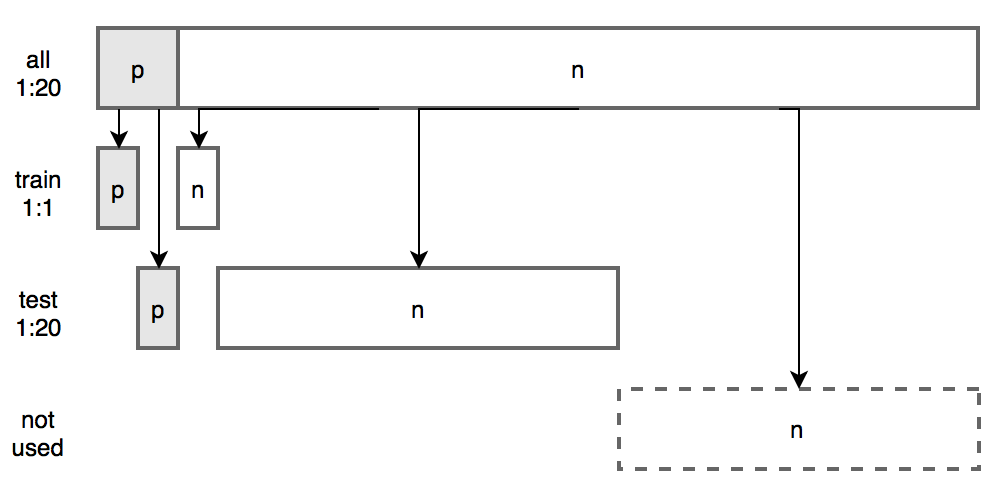}
  \caption{Schematic view of data sampling}
  \label{fig:sampling}
\end{figure}

Initial experiments proved that a training set of $n_{train}' \cup p_{train}'$ containing a ratio of $n_{train}'$ versus $p_{train}'$ similar to $n'$ versus $p'$ in the entire data set, that is 20 non-abused children to one presumably abused child, led to very inferior classifier performance. Therefore, the training set was made up of 50\% $n_{train}'$ and 50\% $p_{train}'$, an equal amount of presumably abused and non-abused children. To ensure a large enough training set whilst retaining enough data for testing, a training set containing $n_{train}'=325$ and $p_{train}'=325$ was initially used. Thus, half of the presumably abused data was used for training. Taking half of the total positive population for training makes it hard to do cross-validation of the results with more than 2 folds, so both the 325 $n_{train}'$ and the 325 $p_{train}'$ were chosen at random from $n'$ and $p'$. This left us with a test set of $n_{test}' \cup p_{test}'$ with $n_{test}' = n'-n_{train}'$ and $p_{test}' = p'-p_{train}'$ for each iteration of cross validation. In order to maintain the ratio between the classes in the test set as it is present in the whole data set, we used $n_{test}' = 0.5 \times (n'-n_{train}')$. Thus, we had 20 non-mistreated children for every presumably abused child, in the test set. We did this as it is important to have the same ratio of p versus n in the test data as in the whole data set, to make sure that the performance metrics are representative for a real life application of the model. \autoref{fig:sampling} provides a visual overview of this sampling method, described as random undersampling \citep{he_learning_2009}. 

\subsection{Benchmark performance}
We first create a benchmark to compare our models with. For this we use a standard approach to text classification with properties as described in \autoref{tab:setup_benchmark_zsl}.

\begin{table}[h]
\caption{Setup of benchmark classifier for ZSL}
\label{tab:setup_benchmark_zsl}       
\def\arraystretch{1}
\begin{tabularx}{\textwidth}{Xl}
\hline\noalign{\smallskip}
\textbf{Setting} & \textbf{Value}  \\
\noalign{\smallskip}\hline\noalign{\smallskip}
Input data & Processed SOCs for age interval 0-4 \\
Features & Count for 100 most common words \\
Classifier & Random Forest (n estimators = 100) \\
\noalign{\smallskip}\hline
\end{tabularx}
\end{table}

\autoref{tab:results_benchmark_zsl} shows the mean performance scores for the model on the test set, using 10-fold cross validation.

\begin{table}[h]
  \centering
  \def\arraystretch{1}
   \caption{Benchmark performance scores for ZSL classification}
  \begin{tabularx}{0.5\textwidth}{Xl}
    \hline\noalign{\smallskip}
    \textbf{Metric} & \textbf{Value} \\
    \noalign{\smallskip}\hline\noalign{\smallskip}
    Precision   & 0.1617  \\
    Accuracy    & 0.8050  \\
    Balanced accuracy & 0.7949 \\
    Recall      & 0.7837  \\
    F1          & 0.2679  \\
    AUC         & 0.8738  \\
    \noalign{\smallskip}\hline
  \end{tabularx}
 
  \label{tab:results_benchmark_zsl}
\end{table}

The algorithm scores quite well on accuracy and recall, but has a low precision score. With the classes in the test data being imbalanced, this implies the presence of many FP instances. This is confirmed by a typical confusion matrix from the test, which in vector form is (TP = 276, FP = 1439, TN = 5529, FN = 56), where for every TP there are about 5 FPs. Nonetheless, we obtain an AUC of 0.87, that already shows promise.


\subsection{Feature building}

The most important features for this bag-of-words approach are the occurrences of words in the texts. A common way to decrease the dimensions of the feature vector is to apply univariate statistical tests to select a top number of features, like the ANOVA or $\chi^{2}$ test. In \autoref{sec:modeltuning} these tests are applied with various parameters. 

In addition to the words as features, we derived some other features from the data given by the JGZ. We appended these to the word-features and tested them for relevance. The following features passed our statistical test:
\begin{itemize}
\item Average amount of characters per consult
\item The most frequently visited JGZ location
\end{itemize}
While, the following features proved not to be relevant include:
\begin{itemize}
\item Lexical diversity
\item Count of family relations per type
\item Gender
\end{itemize}

The most frequently visited JGZ location is a categorical feature, so separate boolean features per location were created forming 142 sparse columns in the feature vector.


\subsection{Algorithm tuning}
In this section we describe how we tuned, analysed and compared the algorithms.

\label{sec:modeltuning}
\subsection{Naive Bayes}
The Naive Bayes model can be applied in two forms: Bernoulli and multinomial. The input of the multinomial algorithm can be weighted, which is often done using tf-idf weighting. These can be smoothed using Laplace or Lidstone smoothing, to account for features that are found in the test set but not in the training data. In this case, especially when using a feature weighting scheme, smoothing will probably not do much for performance. \autoref{tab:nb_perf} contains the results of the grid search that is used to approximate optimal configuration of the Naive Bayes algorithms. The configurations are coded using \textit{mn} for multinomial, \textit{b} for bernoulli and the feature weights \textit{tf-idf}, \textit{cnt} for count and \textit{bool} for boolean.

\begin{table}[h]
\caption{Performance of NB algorithms}
\begin{tabularx}{\textwidth}{X X X X X X l}
\cline{2-7}
  & \multicolumn{3}{l}{\textbf{Accuracy}} & \multicolumn{3}{l}{\textbf{Recall}} \\ \cline{2-7} 
features & mn-tf-idf    & mn-cnt    & b-bool    & mn-tf-idf    & mn-cnt   & b-bool   \\ \hline
100 & 0.7099 & 0.8878 & 0.7532 & 0.5993 & 0.6511 & 0.7277 \\
500 & 0.7406 & 0.8929 & 0.8392 & 0.6262 & 0.6851 & 0.6652 \\
1000 & 0.7461 & 0.9023 & 0.8452 & 0.6270 & 0.6546 & 0.6844 \\
2000 & 0.7272 & 0.9094 & 0.8697 & 0.6837 & 0.6553 & 0.6511 \\
5000 & 0.5629 & \underline{0.9186} & 0.8913 & 0.8170 & 0.6142 & 0.6149 \\
10000 & 0.3029 & 0.9158 & 0.8998 & \underline{0.9376} & 0.6184 & 0.6355 \\ \hline
\end{tabularx}
  
  \label{tab:nb_perf}
\end{table}

\begin{figure}[h]
    \includegraphics[width=1\textwidth]{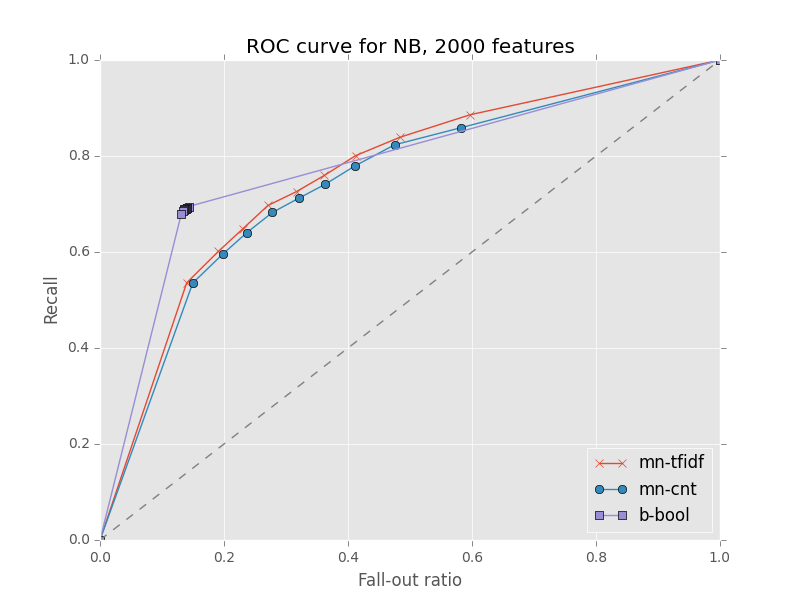}
  \caption{ROC curves for Naive Bayes algorithms}
  \label{fig:roc_nb}
\end{figure}

The outcomes indicate that there is a strong trade-off between the accuracy and recall for the Naive Bayes classifier, and \textit{tf-idf} does not seem to improve the performance. It is unclear whether the multinomial classifiers outperform the Bernoulli variant, but looking at the ROC curves for all three at 2,000 features, it is clear that the "simplest" Bernoulli classifier with boolean input performs the best. Though the precision for this classifier is 0.25, the Bernoulli classifier with Boolean features performs best at an AUC of 0.779. On further testing we found that we reached this score from 50 features and it remained constant up until 10,000 features. We also found that pre-selection of features using $\chi^{2}$ had the most impact, increasing the AUC to 0.817.

\subsection{Random Forest}
We applied a Random Forest (RF) for our decision tree algorithm instead of utilizing a simple decision tree, as it is known to significantly improve the model's performance. Tuning the performance of a Random Forest mainly comes down to selecting the features that are used, the number of trees, and which splitting criterion to employ. We selected the features using ANOVA (shown as \textit{an}) and $\chi^{2}$ tests of relevance, as a RF does not cope well with a very large number of features. For the number of trees, we chose it to be equal to the number of features, as is usually done. For the splitting criterion, we used the default Gini Index splitting criterion in the Scikit-learn package. \autoref{tab:rf_perf} contains the performance results for the models.

\begin{table}[h]
\caption{Performance of Random Forest algorithms}
\begin{tabularx}{\textwidth}{X X X X X X X X l}
\cline{2-9}
  & \multicolumn{4}{l}{\textbf{Accuracy}} & \multicolumn{4}{l}{\textbf{Recall}} \\ \cline{2-9} 
feat & an-cnt & an-tf-idf & $\chi^{2}$-cnt & $\chi^{2}$-tf-idf & an-cnt & an-tf-idf & $\chi^{2}$-cnt & $\chi^{2}$-tf-idf \\ \hline
100 & 0.8357 & 0.8496 & 0.8373 & 0.8566 & 0.7702 & 0.7965 & 0.7546 & 0.7865 \\
200 & 0.8328 & 0.8507 & 0.8476 & 0.8603 & 0.7830 & 0.8121 & 0.7801 & \underline{0.8177} \\
500 & 0.8443 & 0.8651 & 0.8368 & 0.8638 & 0.7652 & 0.8092 & 0.7936 & 0.7993 \\
1000 & 0.8434 & 0.8705 & 0.8441 & 0.8626 & 0.7709 & 0.7830 & 0.7723 & 0.7993 \\
2000 & 0.8501 & 0.8698 & 0.8533 & 0.8669 & 0.7681 & 0.7823 & 0.7865 & 0.7908 \\
5000 & 0.8521 & \underline{0.8768} & 0.8489 & 0.8741 & 0.7816 & 0.7468 & 0.7731 & 0.7901 \\ \hline
\end{tabularx}
\label{tab:rf_perf}
\end{table}

\begin{figure}[h]
    \includegraphics[width=1\textwidth]{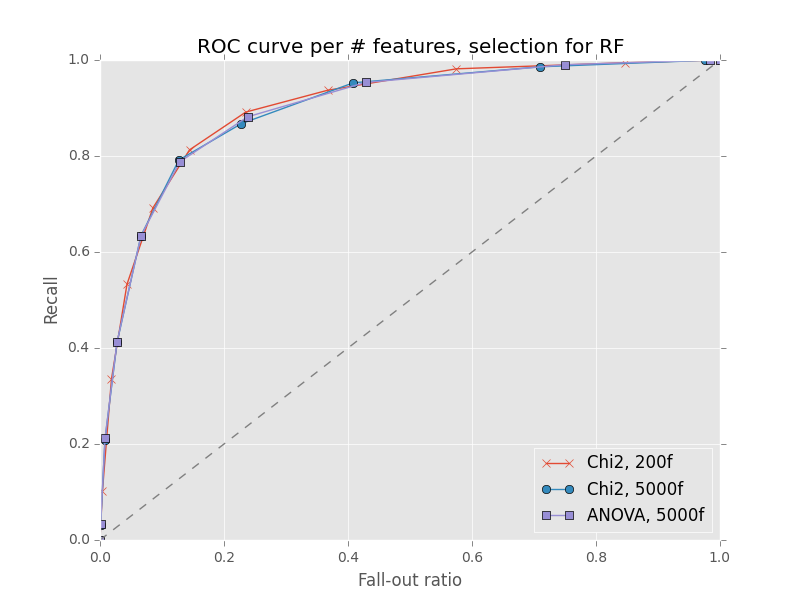}
  \caption{ROC curves for Random Forest algorithms}
  \label{fig:roc_rf}
\end{figure}

From the table it can be easily seen that weighted features result in a higher performance than non-weighted, counted, features. Regarding accuracy, the ANOVA and $\chi^{2}$ feature selection tests scores are similar. For recall, the $\chi^{2}$ test seems to outperform ANOVA when the number of features becomes large. Increasing the number of features results in marginally better accuracy, while the best recall is found by using just 200 features for almost all versions of the algorithm. The precision for this classifier is 0.26. We use an ROC curve to compare the most promising algorithms: \textit{an-tf-idf} and \textit{$\chi^{2}$-tf-idf}. A plot of these curves can be seen in \autoref{fig:roc_rf}.

It is clear that there is not much difference between ROC curves. The computed AUC for $\chi^{2}$ selection with 200 features, 0.903, is slightly higher than the other algorithms at 0.899. The DeltaTF-IDF\footnote{\url{https://github.com/paauw/python-deltatf-idf}} variant of tf-idf leads to an AUC of 0.888, which is lower than the original tf-idf. The BM25\footnote{\url{https://github.com/paauw/python-bm25}} weighting scheme that according to \cite{paltoglou_study_2010} performs best for binary text classification gives an AUC of 0.896. The implementation of these schemes in Python makes them computationally intensive and model run times are 5-10 times longer than the standard tf-idf scheme.

\subsection{SVM}           
SVM is known for the lack of many parameters that can be used for tuning as it does the tuning for a large part internally already. That leaves the user with the type of kernel to be used and the penalty parameter C of the slack error term to tune. We obtained reasonable results with either linear, polynomial or (Gaussian) radial basis function kernels. We used a grid search to approximate the right parameters for the SVM algorithm, as shown in \autoref{tab:svm_perf}.

\begin{table}[h]
\caption{Performance of SVM algorithms}
\begin{tabularx}{\textwidth}{X X X X X X l}
\cline{2-7}
  & \multicolumn{3}{l}{\textbf{Accuracy}} & \multicolumn{3}{l}{\textbf{Recall}} \\ \cline{2-7} 
C & Linear    & Poly    & RBF    & Linear    & Poly   & RBF   \\ \hline
0.2 & 0.8355 & \underline{0.8801} & 0.7150 & 0.7929 & 0.7220 & 0.6348 \\
0.5 & 0.8496 & 0.8596 & 0.7121 & 0.7972 & 0.7170 & 0.6383 \\
0.8 & 0.8491 & 0.8389 & 0.7171 & 0.8121 & 0.7518 & 0.6319 \\
1 	& 0.8425 & 0.8461 & 0.7198 & \underline{0.8248} & 0.7383 & 0.6461 \\
5 	& 0.8168 & 0.8128 & 0.7085 & 0.7731 & 0.7340 & 0.6702 \\
10 	& 0.7982 & 0.7947 & 0.6920 & 0.7766 & 0.7411 & 0.6915 \\ \hline
\end{tabularx}
\label{tab:svm_perf}
\end{table}

\begin{figure}[h]
    \includegraphics[width=1\textwidth]{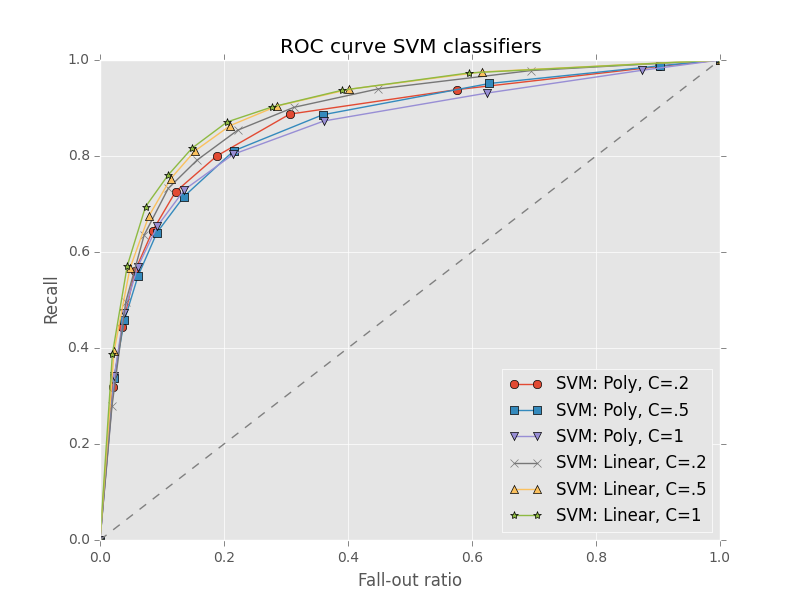}
  \caption{ROC curves for linear and polynomial SVM algorithms}
  \label{fig:roc_svm}
\end{figure}

The precision for this algorithm ranges between 0.21 and 0.23. It can be seen furthermore that the highest F1-score and lowest fall-out rate are achieved when using polynomial C=0.2 algorithm. As recall is an important metric for this model, the polynomial and linear algorithms are compared using an ROC curve for C in [0.2,0.5,1.0]. \autoref{fig:roc_svm} shows the resulting curves.

The ROC curve shows that over the range of C-values, the linear kernel always outperforms the polynomial kernel. Within the linear kernel we find that the differences are marginal and a C-value of 1 can be safely chosen. From here the optimal amount of \textit{tf-idf} weighted textual features can be found at 1,000 with an AUC of 0.906, although using up to 10,000 features can improve recall slightly at the cost of lower accuracy. Pre-selection of features using various methods like ANOVA or $\chi^{2}$ does not effect the model's performance, which is explained by the fact that SVM already selects the most important features into the support vectors. 

We also tested advanced versions of the tf-idf weighting scheme, but they did not result in an improvement of the performance. The DeltaTF-IDF scheme lead to an AUC of only 0.809 and for the BM25 we obtained a AUC of 0.884, both lesser than the basic tf-idf.

\subsection{Additional structured data}
\label{sec:addstruct}
From literature, \autoref{subsec:litabuse}, and the expertise of the JGZ's professional, several structured features were identified that might further enhance the model's performance. The JGZ does not have data on all indicators described in literature, but demographics and a number of other relevant features are available. \autoref{tab:structfeat} contains the data made available by the liaison officer for child abuse. To see whether features built from this data were distinctive and contributed to the model, we applied a $\chi^{2}$ test with respect to the presumed child abuse variable. \autoref{tab:structfeat} shows only the outcomes with $p-values < 0.05$.

Most classifiers can only deal with numeric features and some features are categorical, like the countries of birth. We converted these to sparse matrices with columns enumerating the categorical values. These are so many, that they have not been included in \autoref{tab:structfeat}. Some birth countries correlate significantly  with the presumed child abuse variable, but mainly due to the low volume per country. 

\begin{table}[!htbp]
  \centering
  \def\arraystretch{1}
  \caption{Structured data feature distinctiveness}
  \begin{tabularx}{\textwidth}{Xll}
    \hline\noalign{\smallskip}
    \textbf{Feature} & \textbf{$\chi^{2}$} \\
    \noalign{\smallskip}\hline\noalign{\smallskip}
    JGZ location most visited\textsuperscript{*} & multiple \\
    Birth country of child and both parents\textsuperscript{*} & multiple \\
    Average characters per text & 10373.144  \\
    Age of the mother at child birth & 36.521  \\
    Special consultations on skin issues\textsuperscript{***} & - \\
    Dramatic event\textsuperscript{***} & 99.566  \\
    (Semi)permanent medical condition\textsuperscript{***} & - \\
    General health and disease\textsuperscript{***} & 33.080  \\
    Women's genitalia\textsuperscript{***} & - \\
    Mother's health\textsuperscript{***} & 54.365  \\
    Micturition / defecation\textsuperscript{***} & 6.126  \\
    "Samen Starten weging"\textsuperscript{***} & 296.235  \\
    "Triple-P"\textsuperscript{**} & \\
    \quad 1st contact & - \\
    \quad 2nd contact & - \\
    Burden vs. Carrying\textsuperscript{***} & \\
    \quad Family & 104.438  \\
    \quad Child & 18.050  \\
    \quad Environment & 45.623  \\
    \quad Parents & 90.844  \\
    Dental care\textsuperscript{***} & 5.493  \\
    Overweight\textsuperscript{***} & 9.465  \\
    General care received\textsuperscript{**} & 402.725  \\
    GGD care received\textsuperscript{**} & 111.810  \\
    \noalign{\smallskip}\hline
    \multicolumn{2}{l}{\textsuperscript{*}\footnotesize{Categorical variable}\newline\textsuperscript{**}\footnotesize{Count of occurrences}\newline\textsuperscript{***}\footnotesize{Count of occurrences with findings}} 
  \end{tabularx}
 \label{tab:structfeat}
\end{table}

For this research, there are two interesting applications of this structured data: testing whether a model based on structured data outperforms the model based on unstructured data, and then comparing it with a model build on both structured and unstructured data. 

\subsection{Unstructured versus structured data}

To test whether it makes sense to use the unstructured data  for prediction of presumed abuse instead of using the structured data, we built a classifier based solely on the structured data. We used all the variables stated in \autoref{tab:structfeat} with a Random Forest classifier, that resulted in the performance outcomes in \autoref{tab:performance_struc}. The model's performance is not as good as the model based on unstructured data with respect to recall and accuracy, but the AUC is only slightly worse. Note that the model based solely on structured data makes indirect use of the unstructured data, by using the average count of characters as a feature in this model. Removing this feature only marginally decreases performance, indicating that the feature is mainly important when it can interact with textual features in text mining.

\begin{table}[!htbp]
  \centering
  \def\arraystretch{1}
  \caption{Performance scores for structured classifier}
  \begin{tabularx}{\textwidth}{XXXXXX}
    \hline\noalign{\smallskip}
     & Algorithm & Precision & Accuracy & Recall & AUC \\
    \noalign{\smallskip}\hline\noalign{\smallskip}
    Structured  & RF        & 0.185 & 0.828 & 0.817 & 0.892 \\
    Combined    & SVM       & 0.200 & 0.839 & 0.844 & 0.909 \\
    Ensemble    & RF + RF   & 0.187 & 0.822 & 0.870 & 0.914 \\
    \noalign{\smallskip}\hline
  \end{tabularx}
  \label{tab:performance_struc}
\end{table}

\subsection{Structured and unstructured ensemble}
While a model based solely on structured data did not challenge the model based on unstructured data, a model based on both might perform better than either one individually. In order to to implement the model for the JGZ's daily operations, we need to make sure the model is as good as it can be. Either one model can be built incorporating both forms of data as features, or an ensemble method can be constructed of two models with balanced voting on the outcome.  

The performance of the combined and ensemble classifiers is shown in \autoref{tab:performance_struc}. For the ensemble method a Random Forest algorithm is used for the structured data part and both an SVM and RF were tested for the unstructured part. Since the SVM scored slightly worse (AUC of 0.911) than the RF, the latter was put in the table. The voting is implemented by calculating the average chance of assignment to the positive class between the two classifiers.

\subsection{Acceptance}
\label{subsec:expert_validation}

Although not statistically relevant for improvement of the model, a number of randomly selected model predictions were evaluated by the liaison officer for child abuse. This staff member of the GGD creates the policy related to the signaling and registration of child abuse, and audits the other pediatricians on the subject. The goal of this review was to raise acceptance for the model's performance with the client organization. For the pediatricians, the false positives and false negatives are most interesting: these are the children that might not have been registered correctly. At random, 20 false positive files and 9 false negative files were selected, and reviewed by the liaison officer. She came to the conclusion that:

\begin{itemize}
\item In 13 out of 20 false positive cases presumed abuse should have been registered
\item In 5 out of 20 false positive cases there was reason for doubt, but she would have registered presumed abuse
\item In 2 out of 20 false positive cases there was no presumed abuse: one case of autism and one of severe speech delay
\item In 5 out of 9 false negative cases there was indeed presumed abuse, some cases being quite specific, but also a case of domestic violence
\item In 4 out of 9 false negative cases abuse had wrongly been presumed
\end{itemize}

Upon checking the actions that had been taken by the pediatricians, the liaison officer concluded that in many cases action had been taken accordingly but registration was flawed. This indicates that the problem is mainly with registration of presumed abuse and with ambiguity of the definition of abuse among the pediatricians. The review by the liaison officer proved for the JGZ that the model performs good enough to be useful in day to day operations.

\subsection{Implementation}
\label{subsec:res_implementation}
One of our goals in this research was to have JGZ's pediatricians work with our model to aid their day to day practice. Our model therefore had to be implemented as a decision support system into the existing software used during the child consults. Directly after writing a new SOC, the data could be sent to an API and the classification could then be presented to the user in the form of a warning whenever abuse is presumed. The following action by the pediatrician could be recorded as feedback on the classification by the model. Using the feedback on these classifications, a training set containing less noise can be constructed on-the-fly, possibly improving the model's performance even further. 


\section{Discussion}

From the analysis it is evident that adding meta-features to the textual features improves the model's performance. Next, the best performances are attained when using a boosted Decision Tree algorithm like Random Forest (RF), or when using Support Vector Machine (SVM), that outperforms Naive Bayes (NB) mainly on recall. In order for the algorithms to deliver competitive results, we pre-selected a limited number of features using a statistical test, with $\chi^{2}$ and ANOVA being equally good tests. This was not needed for SVM and might have even made the model performance worse; as SVM can deal with a very large number of features. Furthermore, the features needed to be weighted in any case using a tf-idf variant. We showed that in our case, a more advanced weighting schemes like DeltaTFIDF and BM25 did not outperform the standard tf-idf scheme.

The Area-Under-Curve (AUC) scores for the top RF and SVM classifiers were similar, but indicated that a tuned SVM algorithm performed the best for the prediction of abuse from the unstructured data. This is in line with most text mining literature, that also propose SVM as the best choice algorithm. The SVM algorithm using a linear kernel with C=1 and 1000 tf-idf weighted features performed the best for predicting child abuse from the unstructured data. The computed AUC was 0.906 with an accuracy of 0.843 and recall of 0.825. 

A classifier based solely on structured data did not outperform the SVM classifier based on unstructured data. Combining the structured and unstructured data did however outperform the SVM classifier based solely on unstructured data. The best performance was attained when combining the two classifiers for unstructured and structured data into an ensemble method with an AUC of 0.914, an accuracy of 0.822 and a recall of 0.870.

\subsection{Qualitative evaluation and Raising acceptance}
\label{sec:acceptance}
One of the challenges of our implementation at the JGZ was having a prototype based on our prediction model accepted by the employees at JGZ. We therefore decided to get a small random sample of files and their respective model predictions reviewed by the liaison officer for child abuse at JGZ. This is not relevant for validation of the model, as the model already contains the judgment of around 500 professionals, and validating outcomes by one professional would not be statistically sound. Using such a predictive model in practice however, is a big step towards acceptance of the results for a successful implementation. 29 children's files where selected at random from the group of false positives (FP) and false negatives (FN), 20 false positives and 9 false negatives. More FP files were selected, as the FP files were most interesting from the perspective of JGZ's work and time that could be spent by the liaison officer was limited. 

The liaison officer for child abuse reviewed the files to see whether she agreed with the model. She came to the following conclusions:

\begin{itemize}
\item In 13 out of 20 false positive cases presumed abuse should have been registered
\item In 5 out of 20 false positive cases there was reason for doubt, but she would have registered presumed abuse
\item In 2 out of 20 false positive cases there was no presumed abuse: one case of autism and one of severe speech delay
\item In 5 out of 9 false negative cases there was indeed presumed abuse, some cases being quite specific, but also a case of domestic violence
\item In 4 out of 9 false negative cases abuse had wrongly been presumed
\end{itemize}

Upon checking the actions that had been taken by the paediatricians, the liaison officer concluded that in many cases action had been taken accordingly but registration was flawed. This indicates that the problem is mainly with registration of presumed abuse and with ambiguity of the definition of abuse among the paediatricians. 

\subsection{Implementation}
\label{subsec:implementation}
\cite{bellazzi_predictive_2008} state that the project is not finished with a good model, one should ensure that the decision support system is properly implemented. This was also the goal of this research: to support the medical staff with the implemented model. To create impact with such an implementation, it was essential to let the client at the JGZ decide on the most appropriate form of implementation. 


Together with the JGZ and the developer of their client management tool, it was decided to make the functionality of the prediction model available through an API (Application Programming Interface). The API would be called from within their existing software upon closing a child's file. A prediction will then be made based on the entire file, including anything that had just been appended to it. The user is warned via a popup whenever a child's file is marked by the model as a case of possible abuse. This should urge the professional to take action and correctly register it if needed.

An important part of the implementation was the option for a professional to provide the model with feedback on its performance. This feedback would be then used in the next learning cycle of the model and would thus improve the model's performance in the long run.

\subsection{RESTful API}
We developed an API following the REST (Representational State Transfer) architecture style. In this way, the functionality of the model could be easily incorporated in any data management system that a client uses, while not requiring a direct connection with the client's database. For the prototype, we wrapped Python-based trained models in an API using the Flask\footnote{\url{http://flask.pocoo.org}} framework. There were two endpoints to this API: \textit{/predict} that requires the POST method to provide the features and \textit{/feedback} that processes POST-ed feedback on predictions.

The first endpoint required the features of the model as input, that is: all SOCs from the file and all structured variables used by the structured classifier in the ensemble. The API saved the data to a database which also contained training and test data. Next, the model made a prediction for the child and returned that to the caller, together with the ID of the newly created database record. This ID could be used to later provide feedback on the prediction. In a given time period (like every night), the model would retrain itself using both the training and the feedback data. \autoref{fig:api_proc} provides a schematic overview and  illustrates these processes of retrieving a prediction, sending feedback and re-training the model.

Our API ran in a secure environment and was reachable via an SSL (Secure Socket Layer) TLS connection. All the data was transferred and saved exclusively in an anonymous form, due to the sensitive nature of the data. We have implemented the API along with these security measures in the JGZ's client management system. On an average day, the medical professionals spread over 27 JGZ locations in a region of The Netherlands  consult around 650 children between 0-4 (2014), equaling to around 650 API calls per day. 

\section{Conclusions}
\label{sec:conclusions}

In this article we proposed a decision support system for identifying child abuse based on structural and free-text data. A systematic review of machine learning methods showed that the free-text data can indeed be used to signal presumed abuse. Both structured and unstructured data contain meaningful patterns that we used to create  Random Forest, and support vector machine models. We achieved the highest score on the AUC-metric, which we identified as the most appropriate evaluation metric, by using an ensemble classifier combining the structured and unstructured data. 

The performance of the decision support system was not only evaluated mathematically, but also by comparing its classifications with those made by the liaison officer from JGZ, an expert on child abuse. The high degree of agreement between this expert and our ensemble classifier lead to a wide acceptance of the proposed decision support system among the end users from the Dutch youth health care agency (JGZ). 

We can therefore summarize our contributions in this article as follows:
\begin{itemize}
\item we have shown that free-text data by pediatricians contains patterns to label child abuse,
\item we have compared several machine learning methods for their effectiveness in this task,  
\item we have evaluated some of these methods through health care professionals, and
\item we then also describe our implementation of the prediction model through an API.
\end{itemize}

This research has shown that utilizing machine learning techniques for children's health-related decision support is both feasible and beneficial. There are many ways in which this research can be further extended. Future work can focus on one of the following directions:
\begin{itemize}
\item Including more data from other relevant agencies, e.g. schools.
\item Weighting evidence according to its temporal distance from the present moment.
\item Extend the models for other threats to children's health.
\item Evaluate the long-term effects of the automated identification of child abuse.
\end{itemize}

Our findings have the potential to improve the correct registration of child abuse, which is an important step to its prevention and the reduction of its effects. 

\appendix

\bibliography{sample}
\bibliographystyle{apalike}

\label{appendix-A}
\section{Data Clusters}
\label{app:dataClusters}
We ran the algorithm for 10 clusters, and obtained the terms shown in \autoref{tab:cluster10}.

\begin{table}[!htbp]
\caption{Cluster features for 10-means clustering}
\label{tab:cluster10}
  \def\arraystretch{1}
\begin{tabularx}{\textwidth}{llX}
\hline\noalign{\smallskip}
Cluster & Children & Features  \\
\noalign{\smallskip}\hline\noalign{\smallskip}
0 & 1139 & verhuiz, babi, pgo, werk, kindj, gezin, gebor, meisj, gegroeid, terug, vacc, beval, last, consult, groeit \\
1 & 1040 & afsprak, kinder, contact, gebeld, nvzb, nieuw, thui, hulp, woont, vve, gezin, moedervanmoed, meisj, schol, gesprok \\
2 & 623 & fysiotherapeut, vkh, re, meisj, verwez, hoofd, kdv, bespr, recht, li, folder, ka, gegroeid, beter, eten \\
3 & 775 & ka, co, control, afsprak, meisj, oc, fysiotherapeut, beter, oogart, kdv, ziekenhui, verwez, lengt, vacc, kinder \\
4 & 1028 & nederland, vve, spreekt, meisj, ned, vacc, ooi, prat, tal, thui, woord, afsprak, kinder, allen, vakanti \\
5 & 2000 & meisj, vlot, mooi, groeit, bespr, slaapt, kdv, gegroeid, gezond, eten, nacht, werk, drinkt, som, vacc \\
6 & 1760 & mannetj, ventj, jongetj, jong, mooi, bespr, kdv, groeit, slaapt, verkoud, vlot, gegroeid, vacc, som, gezond \\
7 & 1311 & slap, nacht, slaapt, kdv, eten, wakker, beter, drinkt, fle, som, overdag, meisj, vak, bed, groeit \\
8 & 854 & logopedi, vve, woord, logo, prev, afsprak, nederland, maakt, prat, ooi, contact, verwez, thui, gehor, tal \\
9 & 1301 & meisj, vacc, zuigel, ooi, lengt, leeftijd, bmi, gezond, kdv, gegroeid, vov, klein, melk, apk, eten \\
\noalign{\smallskip}\hline
\end{tabularx}
\end{table}

Here we see clusters for girls (5, 9) and boys (6). Other themes that are differentiated on include moving (0), making/forgetting appointments (1), the physiotherapist (2, 3), speaking Dutch (4), sleeping issues (7) and speech issues (8). 

\section{API Implementation}
\label{app:ApiImpl}
\begin{figure}[!htbp]
    \includegraphics[width=1\textwidth, scale=0.7]{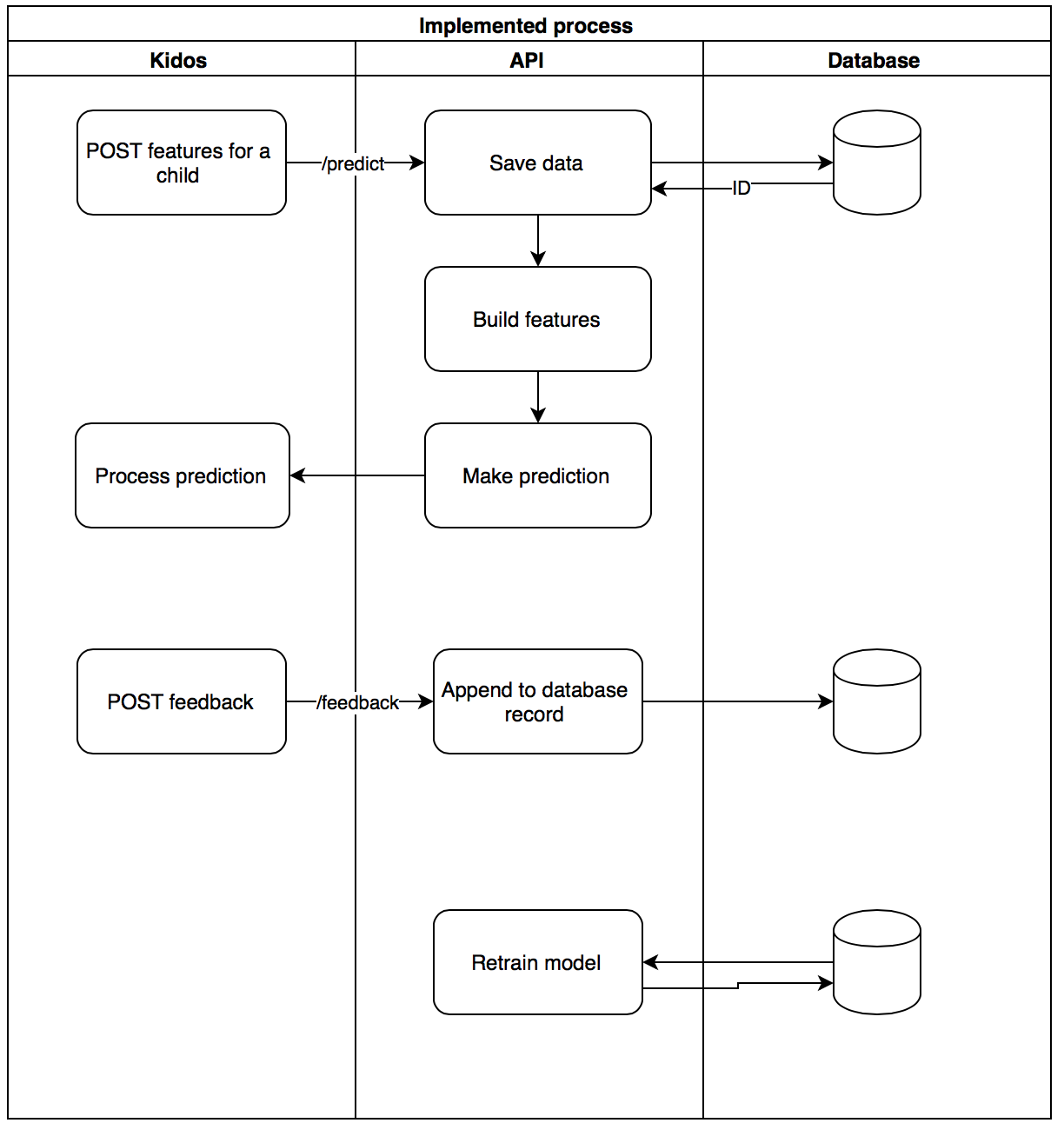}
  \caption{Schematic view of the API implementation}
  \label{fig:api_proc}
\end{figure}

\FloatBarrier
\section{Initial layout}
\label{app:initiallayout}

The data set from the GGD is split into several files, each having a specific subject. An overview of this data is included in \autoref{tab:datalayoutbefore}.

\begin{table}[!htbp]
\def\arraystretch{0.8}
\caption{Layout of the GGD data set}
\begin{tabularx}{\textwidth}{lll}
\hline\noalign{\smallskip}
File & Column & Type  \\
\noalign{\smallskip}\hline\noalign{\smallskip}
Conclusions & Person number & integer \\
& Birth date & date \\
& JGZ location & text \\
& Action type & text \\
& Observation date & date \\
& Conclusion & text \\
Family relations & Person number & integer \\
& Child birth date & date \\
& Relation type & text \\
& Person number & integer \\
& Child relation birth date & date \\
BMI & Person number & integer \\
& Birth date & date \\
& Sex & text \\
& Action type & text \\
& Length & float \\
& Weight & float \\
& BMI date & date \\
& BMI age & float \\
& BMI & float \\
Worries ZSL & Person number & integer \\
& Birth date & date \\
& JGZ location & text \\
& Action type & text \\
& Observation type & text \\
& Value & text \\
Findings ZSL & Person number & integer \\
& Birth date & date \\
& Action type & text \\
& Finding date & date \\
& Finding type & text \\
& Finding & text \\
Actions ZSL & Person number & integer \\
& Birth date & date \\
& JGZ location & text \\
& Action type & text \\
& Observation type & text \\
& Action & text \\
Attention child & Person number & integer \\
& Birth date & date \\
& Attention & boolean \\
\noalign{\smallskip}\hline
\end{tabularx}

\label{tab:datalayoutbefore}       
\end{table}

\end{document}